\documentclass[a4paper]{raa}

\usepackage{graphicx,times}
\usepackage{natbib}
\usepackage{amssymb,amsmath}

\bibpunct{(}{)}{;}{a}{}{,}

\usepackage[a4paper=true,dvipdfm=true,pagebackref=true]{hyperref}
\hypersetup{pdftitle={Performance analysis of \textit{N}-body codes}, pdfauthor={S. Huang, R. Spurzem \& P. Berczik}} 
\hypersetup{colorlinks = true, linkcolor = green, anchorcolor = red, citecolor = blue, filecolor = red, pagecolor = red, urlcolor = red}

\newcommand{\nbody}{\textit{N}-body\ }
\newcommand{\mynb}{\texttt{NBODY6++}\ }
\newcommand{\mybs}{\texttt{Bonsai}\ }

\begin{document}

   \title{ Performance analysis of parallel gravitational \textit{N}-body codes on large GPU cluster
   }

   \volnopage{Vol.0 (20xx) No.0, 000--000}
   \setcounter{page}{1}
   
   \author{ Siyi Huang\inst{1}, Rainer Spurzem\inst{1,2,4}, Peter Berczik\inst{1,3,4}
   }

   \institute{ National Astronomical Observatories and Key Laboratory of Computational Astrophysics, Chinese Academy of Sciences, Beijing 100012, China; {\it huang41@nao.cas.cn}
               \and
               Kavli Institute for Astronomy and Astrophysics, Peking University, Beijing 100871, China
               \and
               Main Astronomical Observatory, Ukrainian National Academy of Sciences, 03680 Kiev, Ukraine
               \and
               Astronomisches Rechen-Institut, Zentrum f{\"u}r Astronomie, Universit{\"a}t Heidelberg, 69120 Heidelberg, Germany
   }
   
   \abstract{We compare the performance of two very different parallel gravitational \textit{N}-body codes for astrophysical simulations on large GPU clusters, both pioneer in their own fields as well as in certain mutual scales - {\tt NBODY6++} and {\tt Bonsai}. We carry out the benchmark of the two codes by analyzing their performance, accuracy and efficiency through the modeling of structure decomposition and timing measurements. We find that both codes are heavily optimized to leverage the computational potential of GPUs as their performance has approached half of the maximum single precision performance of the underlying GPU cards. With such performance we predict that a speed-up of $200-300$ can be achieved when up to 1k processors and GPUs are employed simultaneously. We discuss the quantitative information about comparisons of two codes, finding that in the same cases {\tt Bonsai} adopts larger time steps as well as relative energy errors than {\tt NBODY6++}, typically ranging from $10-50$ times larger, depending on the chosen parameters of the codes. While the two codes are built for different astrophysical applications, in specified conditions they may overlap in performance at certain physical scale, and thus allowing the user to choose from either one with finetuned parameters accordingly.     
   }
   
   \authorrunning{S. Huang, R. Spurzem \& P. Berczik}
   \titlerunning{Performance analysis of \nbody codes}
   \maketitle

%

\section{INTRODUCTION}
\label{sec:introduction}
Algorithms for gravitational \nbody simulations, which are widely used tools in astrophysics nowadays, have mainly evolved into two categories over the passed decades. 
Traditionally, by computing the pair-wise force among particles, the direct summation method has been employed as the core idea of the so-called ``direct $N$-body code''. High accuracy can be archived by choosing smaller time steps, with higher computational costs. Some best-known examples are the \texttt{NBODY} series codes developed by \citet{Aarseth+1999} and the \texttt{Starlab} environment developed by Hut, McMillan, Makino, Portegies Zwart, etc \citep{Hut+2003}. 
Alternatively, the force calculation can be approximated with certain assumptions. In the late 1960s, some approximation algorithm such as tree-code or mesh-code are developed in attempt to reduce the computational complexity so that larger simulations can be scaled on limited hardware with acceptable time. One of the most prominent approximated algorithm, namely Barnes-Hut tree \citep{barnes+hut+1986}, also have many implementations such as \texttt{GADGET} developed by \citet{Springel+2005} and \texttt{PEPC} developed by Gibbon, Winkel and collaborators \citep{Winkel+etal+2012} .

Since direct summation methods use an all-to-all particle force direct summation method they have a raw computational complexity $\mathcal{O}(N^2)$. Additional algorithms are developed to cut the absolute wall-clock time down in spite of the prohibitive asymptotic complexity. On the other side approximate schemes reduce complexities to $\mathcal{O}(N \log N)$ or even $\mathcal{O}(N)$ thanks to the approximate treatments in force computations and some special structures such as octree or grids.
However, such intelligent approximative algorithms may not be very suitable for the simulation of certain astronomical systems such as dense star clusters, e.g. globular clusters or nuclear star clusters with or without central massive black holes. This is because in these systems two-body relaxation is important, which can only be correctly modeled by following pairwise particle interactions with high precision at large distances. They require the use of direct summation methods, which has so far great difficulties reaching even one million particles (but see \citealt{Wang+etal+2015}). As such, the only practical approach at the present to handle simulations with ultra-high particle numbers (e.g. cosmological structure formation) is through the employment of approximate methods, despite their lack of resolution at small scales \citep{Shin+etal+2014, Genel+etal+2014, Vogelsberger+etal+2014}.

Consequently, parallel technologies are applied for \nbody simulations as a proper solution. With the help of parallel hardware, simulations could be accelerated multifold in accordance with the numbers of invoking processors theoretically.
In practice parallel schemes were implemented, as well as performance analysis of parallel \nbody codes on supercomputers or distributed systems, such as the work by \citet{Gualandris+etal+2007}.
Supercomputer clusters were used for the parallelization of \nbody simulations, special devices were also added in order to process the hot computation sections. 
In the beginning, special-purpose architectures called GRAPE series \citep{Makino+etal+2003} are designed for \nbody simulations exclusively, which achieved speed-ups by putting the whole force calculations into the hardware that placing many pipelines on one chip, detailed performance of GRAPE measured by \citet{Harfst+etal+2007}. In recent years, with the rapid development of hardware manufacturing technique the GPU (Graphics Processing Unit) as general-purpose devices are more and more used and acted the same role as GRAPE. Now architectures consisted of many processors and equipped with corresponding GPUs are prevalent solutions in \nbody simulations.

Parallel \nbody simulation software running partially or even entirely on GPUs were developed subsequently \citep{Berczik+etal+2011, Berczik+etal+2013, Spurzem+etal+2012,   Bedorf+etal+2012a, Bedorf+etal+2012b}, while in practice applications the performance would not measure up to ideal speed-up because of some inevitable serial code in the code structure. The actual speed-up is limited by sequential fractions in codes and not directly proportional to the number of processor cores, the theoretical maximum value could be predicted by \emph{Amdahl's law}  \citep{Amdahl+1967}. What's more, this peak value is unapproachable on account of communication overhead between multiple processors.
The effectiveness of either parallelization or GPU acceleration introduced in \nbody software is not intuitive but interested. 

In this paper, we focus on the performance analysis of two kinds of \nbody software, direct \nbody code \mynb and tree-code \mybs, which both can be executed in parallel and accelerated by GPUs. Section~\ref{sec:condition} describes an overview of the software and hardware we used. Section~\ref{sect:performance} describes the performance models used to analyse complicated \nbody codes, provides the detailed measurements, performance results and reasonable predictions. Section~\ref{sect:discussion} describes the performance comparisons and analysis, then makes a conclusion which gives us a better reference in the choice of opportune scheme of software type and hardware scale in \nbody simulations.

\section{SOFTWARE AND HARDWARE}
\label{sec:condition}

\subsection{Direct \nbody implementation: \mynb}

In this section we provide a brief description of \mynb, which we used for the performance analysis of direct \nbody code.

\mynb developed by Rainer Spurzem is a parallel version of \texttt{NBODY6} \citep{Spurzem+1999, Khalisi+etal+2003, Spurzem+etal+2008}.
The standard \texttt{NBODY6} is the $6^{\mathrm{th}}$ generation of \texttt{NBODY} code initiated by Sverre Aarseth, who has a lifelong dedication to the development of the family of \texttt{NBODY} series codes \citep{Aarseth+1999}. The first code \texttt{NBODY1} was a basic direct \nbody code with individual time steps. Ahmad-Cohen neighbour scheme \citep{Ahmad+Cohen+1973} was used in \texttt{NBODY2} and \texttt{NBODY5} made it possible to treat larger systems. Kustaanheimo-Stiefel (1965) two-body regularization and chain regularization were applied in \texttt{NBODY3} and \texttt{NBODY5} to deal with close encounters. By the time \texttt{NBODY6} had developed, the code included both neighbour scheme and regularizations, as well as applied with Hermite scheme integration method combined with hierarchical block time steps.

\mynb is a descendant of the standard \texttt{NBODY6}, it kept those features of its predecessor mentioned above as well as increased the efficiency by redesigning the algorithms suitable for parallel hardware.
\mynb used the SPMD (Single Program Multiple Data) scheme to achieve parallelism, in this mode multiple autonomous processors simultaneously start with chunked local data and then communicate with each other through the \emph{copy algorithm} \citep{Makino+2002, Dorband+etal+2003}, which is a parallelized algorithm assumes that each processor has a local copy of the whole system and every processor handles the subgroup of data itself then broadcasts the new data to all the other processors immediately. The parallelization scheme of \mynb is implemented with the standard MPI library package.

The most major improvement of \mynb is the parallelization of regular and irregular force computations, which were special concepts introduced from Ahmad-Cohen neighbour scheme that divided the full force of each particle involved by other all particles into two parts: one part called irregular force that has a frequent but short time steps for interactions with adjacent particles, another one called regular force which has a longer time steps for full interactions. By assigning the most expensive overhead sections to multiple processors \mynb achieved the expected efficiency. Moreover, the heavier regular force computation component was adapted for GPU acceleration using CUDA. The performance of parallel accelerations showed in Section~\ref{sec:nbody6++ptitle}.

\subsection{\nbody tree-code implementation: \mybs}

In this section we provide a brief description of \mybs, which we used for the performance analysis of \nbody tree-code.

Tree-code algorithm as a widely used method nowadays for \nbody simulations is originally introduced by \citet{barnes+hut+1986}. This algorithm reduces the computational complexity of \nbody simulation from $\mathcal{O}(N^2)$ to $\mathcal{O}(N \log N)$, therefore improves the simulating scale compared to the brute force methods. Here \mybs, a parallel GPU tree-code implementation developed by Jeroen B{\'e}dorf, Evghenii Gaburov and Simon Portegies Zwart \citep{Bedorf+etal+2012a, Bedorf+etal+2012b}, is a suitable representative of gravitational \nbody tree-code in recent years.

Certain schemes are introduced in \mybs to ensure the high efficiency of the code \citep{Bedorf+etal+2012a}. 
A sparse octree is used as the data structures, which means the structure is the three-dimension extension of a binary tree where tree-cells are not complete and equal, it is based on the underlying particle distribution. The tree is constructed layer-by-layer from top to bottom, and inverted the direction in traverse process. Tree-cell properties are updated during the steps, and the integration of the simulation are advanced.	
The depth of tree traversing affects both accuracy and time consumption crucially, which is determined by the multipole acceptance criterion (MAC) in the tree-code. The criterion is described as follows,
\begin{equation} \label{eqmac}
  d > \frac{l}{\theta} + \delta
\end{equation}
where $d$ is the smallest distance between a group and the cell's center-of-mass, $l$ is the length of the cell, $\delta$ is the distance between the cell’s center-of-geometry and the center-of-mass, $\theta$ is an opening angle parameter to control the accuracy. If the inequality is satisfied then traversing process will be interrupted while multipole moment will be used. 

Like other existing \nbody codes, \mybs uses parallel technique to reach large scale or high resolution simulations, and applies GPUs to speed up force computations. In contrast to those GPU tree-code, \mybs executes all parts of the algorithm on GPUs, to avoid the bottlenecks generated from CPU-GPU communications. In Section~\ref{sec:bonsaiptitle} the performance of main parts in the code are presented.

\subsection{Hardware environments and initial conditions}
\label{sec:hardware}

The supercomputer we mainly used for all the tests of both software presented above is an IBM iDataPlex Cluster JUDGE named ``The Milky Way System" partition provided and maintained by J{\"u}lich Supercomputing Centre in Germany, which is a dedicated GPU cluster using 2 Intel Xeon X5650 6-core processors and 2 NVIDIA Tesla M2050/M2070 GPU cards in every node, with 206 compute nodes and 239 Teraflops peak performance in total.

Our performance measurements involved two different kinds of parallel gravitational GPU-accelerated \nbody codes: \mynb and \mybs. The initial conditions of all tests of both codes are consistent with each other, starting with Plummer model and running over standard $1$ \textit{N}-body Time Unit. The number of particles used ranges from $N=2^{13}(8\mathrm{k})$ to $2^{20}(1\mathrm{M})$, doubled the number over the interval successively. There are additional tests using larger particle numbers up to $N=2^{24}(16\mathrm{M})$ in \mybs code runs. The number of processors we chose is the series increasing number $N_p=1, 2, 4, 8, 16, 32$. Other parameters which are necessary but specific only in each code, such as time step factor for regular/irregular force polynomial and desired optimal neighbour number in \mynb, or accuracy control parameter $\theta$ and softening value $\epsilon$ in \mybs, will be described detailed in Section~\ref{sect:performance}.

\section{PERFORMANCE}
\label{sect:performance}

In this section we evaluated the performance of these two GPU-based parallel \nbody simulation codes (i.e. \mynb and \mybs) which we tested mainly on the J{\"u}lich Dedicated GPU Environment described in Section~\ref{sec:hardware}.

In spite of a vast difference between two codes derived from their own fundamental algorithms and specific details, which make it difficult to give a one-to-one comparison, there are some global values providing sufficient information. Timing variables, speed-up and hardware performance indicators like speed and bandwidth were measured below for performance analysis of codes.

\subsection{Performance of \mynb}
\label{sec:nbody6++ptitle}

\begin{figure}
   \centering
   \includegraphics[width=12.0cm, angle=0]{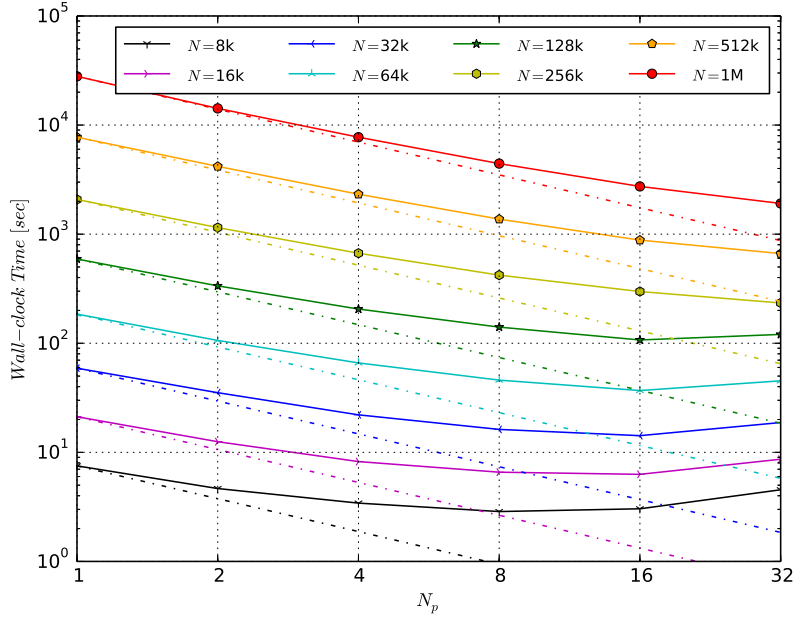}
   \caption{Total wall-clock time ($T_{\mathrm{tot}}$) of \mynb as a function of $N$ \& $N_p$. Solid lines are the measured values of running time, dashed lines are the ideal acceleration by increasing processor numbers. (The unit symbols in the legend have the magnitudes: $1\mathrm{k}=1,024$, $1\mathrm{M}=1\mathrm{k}^2$ and $1\mathrm{G}=1\mathrm{k}^3$, similarly hereinafter.) }
   \label{Fig1}
\end{figure}

\mynb, a parallel direct gravitational \nbody code, is featured with a couple of elegant algorithms and schemes developed and maintained over the past few decades. The procedures of durative ameliorations enabled more realistic size of simulations running in achievable circumstances while increased sophistication as well.
As a consequence we present a performance model for analysing the overall behaviour as well as main components of the code. Through this model we will have a better idea about the performance of a typical direct \nbody code and predictions about the code behaviour further in larger scales.

\subsubsection{Performance model}
\label{sec:nbody6++model}

We measured running time directly for evaluating performance and modelling. 
In \mynb, the total wall-clock time $T_{\mathrm{total}}$ required to advance the simulation for a certain integration interval can be written as
\begin{equation} \label{eqnbt}
  T_{\mathrm{tot}} = T_{\mathrm{force}} + T_{\mathrm{comm}} + T_{\mathrm{host}}
\end{equation}
where $T_{\mathrm{force}} = T_{\mathrm{reg}} + T_{\mathrm{irr}} + T_{\mathrm{pre}}$ is time spend on both host and device involving force calculations, here $T_{\mathrm{reg}}$, $T_{\mathrm{irr}}$ and $T_{\mathrm{pre}}$ are time spend on force computations of regular time step, irregular time step respectively and prediction; 
$T_{\mathrm{comm}} = T_{\mathrm{mov}} + T_{\mathrm{mci}} + T_{\mathrm{mcr}} + T_{\mathrm{syn}}$ is time spend on data moving for parallel components, MPI communications after regular and irregular blocks and synchronizations of processors; 
$T_{\mathrm{host}}$ is time spend on host side which are absolutely sequential runs. 
All of the time variables are measured directly by standard Fortran functions \texttt{ETIME} in sequential mode and \texttt{MPI\_WTIME} in parallel mode. 
The entire descriptions are gathered in the last glossary (Table~\ref{table3}).

According to the decomposition described above we broke down the code structure and measured these main sections which have heavy weights in code. Owing to a large amount of variables and the high complexities some insignificant components in \mynb are not counted in. 
For every parts to be analysed we listed the expected scaling and optimal fitting value in Table~\ref{table1}, which are gotten from the structure of code implementation, chronograph and fitting functions. A python function \texttt{scipy.optimize.curve\_fit} are used to obtain the optimal fitting value, which based on non-linear least squares.

\begin{table}
\bc
\begin{minipage}[]{100mm}
\caption[]{Main components of \mynb \label{table1}}
\end{minipage}
\setlength{\tabcolsep}{1.0pt}
\small
\newcommand{\myboxa}[1]{\raisebox{-2ex}[0pt][0pt]{\shortstack{#1}}}
\newcommand{\myboxb}[1]{\raisebox{-3ex}[0pt][0pt]{\shortstack{#1}}}
\begin{tabular}{ccccc}
\hline
\myboxa{Description} & \myboxb{Timing \\ variable} & \multicolumn{2}{c}{Expected scaling} & \myboxa{Fitting value [sec]} \\
 & & $N$ & $N_p$ & \\
\hline
Regular force computation & $T_{\mathrm{reg}}$ & $\mathcal{O}(N_{\mathrm{reg}}\cdot N)$ & $\mathcal{O}(N_p^{-1})$ & $(2.2\cdot 10^{-9}\cdot N^{2.11}+10.43) \cdot N_p^{-1}$\\
Irregular force computation & $T_{\mathrm{irr}}$ & $\mathcal{O}(N_{\mathrm{irr}}\cdot \langle N_{nb} \rangle)$ & $\mathcal{O}(N_p^{-1})$ & $(3.9\cdot 10^{-7}\cdot N^{1.76}-16.47) \cdot N_p^{-1} $ \\
Prediction & $T_{\mathrm{pre}}$ & $\mathcal{O}(N^{kn_{p}})$ & $\mathcal{O}(N_p^{-kp_{p}})$ & $(1.2\cdot 10^{-6}\cdot N^{1.51}-3.58) \cdot N_p^{-0.5}$ \\
Data moving & $T_{\mathrm{mov}}$ & $\mathcal{O}(N^{kn_{m1}})$ & $\mathcal{O}(1)$ & $2.5\cdot 10^{-6}\cdot N^{1.29}-0.28$ \\
MPI communication (regular) & $T_{\mathrm{mcr}}$ & $\mathcal{O}(N^{kn_{cr}})$ & $\mathcal{O}(kp_{cr}\cdot \frac{N_p-1}{N_p})$ & $(3.3\cdot 10^{-6}\cdot N^{1.18}+0.12)(1.5 \cdot \frac{N_p-1}{N_p})$ \\
MPI communication (irregular) & $T_{\mathrm{mci}}$ & $\mathcal{O}(N^{kn_{ci}})$ & $\mathcal{O}(kp_{ci}\cdot \frac{N_p-1}{N_p})$ & $(3.6\cdot 10^{-7}\cdot N^{1.40}+0.56)(1.5 \cdot \frac{N_p-1}{N_p})$ \\
Synchronization & $T_{\mathrm{syn}}$ & $\mathcal{O}(N^{kn_{s}})$ & $\mathcal{O}(N_p^{kp_{s}})$ & $(4.1\cdot 10^{-8}\cdot N^{1.34}+0.07) \cdot N_p$ \\
Sequential parts on host & $T_{\mathrm{host}}$ & $\mathcal{O}(N^{kn_{h}})$ & $\mathcal{O}(1)$ & $4.4\cdot 10^{-7}\cdot N^{1.49}+1.23$ \\
\hline
\end{tabular}
\ec
\tablecomments{0.8\textwidth}{Detailed descriptions of used symbol gathered in Table~\ref{table3}.}
\end{table}

The conception speed-up is used for evaluating the parallelism of the code. There are a couple of definitions of speed-up with different ranges.
The ideal maximum speed-up $S_i = N_p$ will never be accessible, where $N_p$ is the number of processors used. 
Unreachable as well, but a more reasonable indicator to predict the theoretical maximum speed-up is so-called \emph{Amdahl's law}, which is defined as
\begin{equation} \label{eqsa}
  S_a(N_p) = \frac{T(1)}{T(N_p)} = \frac{1}{(1-X)+\frac{X}{N_p}}
\end{equation}
where $X$ is the fraction of the algorithm that can benefit from parallelization.
In practice there is another experiential speed-up to be measured through timer recording, which is given by
\begin{equation} \label{eqse}
  S_e(N_p) = \frac{T_{\mathrm{tot}}(1)}{T_{\mathrm{tot}}(N_p)}
\end{equation}
where $T_{\mathrm{tot}}(1)$ \& $T_{\mathrm{tot}}(N_p)$ are both the measured values of actual running time.
By combining the fitting values into the speed-up formula we will have a general overall perception of the code, and by which make a prediction accordingly about the code performance in larger scale simulations.

The speed of force calculation is measured by the extent at which program reaches the peak of computing devices. Here in our tests the computing device particularly refers to the NVIDIA Tesla M2050/M2070 GPU cards, which feature up to 1,030 Gigaflops of single precision floating point performance and 515 Gigaflops of double precision floating point performance per card.

\begin{figure}
   \centering
   \includegraphics[width=12.0cm, angle=0]{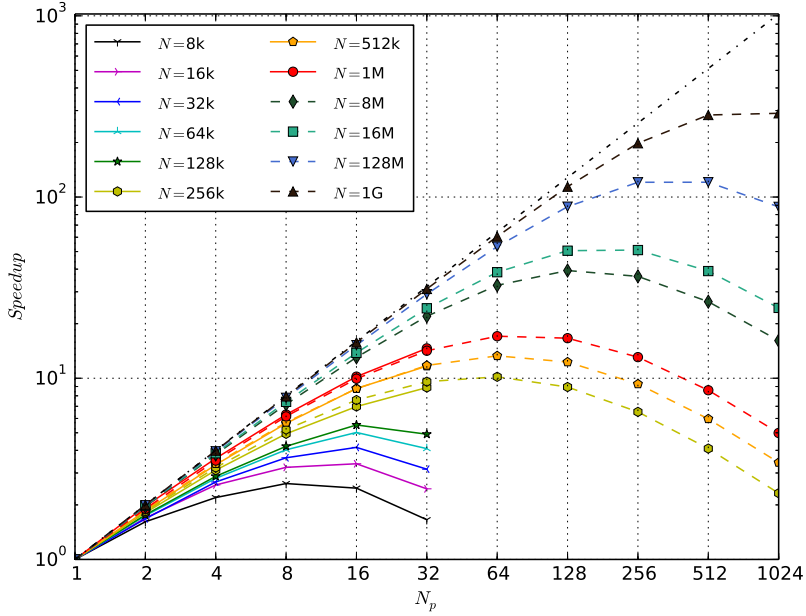}
   \caption{The speed-up ($S$) of \mynb as a function of $N$ \& $N_p$. Solid lines are the measured speed-up ratio between sequential and parallel wall-clock time, dashed lines are the predicted performance of larger scale simulations.}
   \label{Fig2}
\end{figure}

In \mynb, as the total force are divided into two parts we used two speed variables $P_{\mathrm{reg}}$ and $P_{\mathrm{irr}}$ represent regular and irregular force calculating speed, which are written as
\begin{equation} \label{eqnbp}
P_{\mathrm{reg}} = \frac{N_{\mathrm{reg\_tot}}}{T_{\mathrm{reg}}} = \frac{N_{\mathrm{reg}} \cdot N \cdot \gamma_{\mathrm{h4}}}{T_{\mathrm{reg}}}
,\qquad
P_{\mathrm{irr}} = \frac{N_{\mathrm{irr\_tot}}}{T_{\mathrm{irr}}} = \frac{N_{\mathrm{irr}} \cdot \langle N_{\mathrm{nb}} \rangle \cdot \gamma_{\mathrm{h4}}}{T_{\mathrm{irr}}}
\end{equation}
where $N_{\mathrm{[reg|irr]\_tot}}$ is the total floating point operations of regular/irrgular force computations, $N_{\mathrm{[reg|irr]}}$ is the cumulative number of regular/irrgular time steps, $\langle N_{\mathrm{nb}} \rangle$ is the average number of integrated ``neighbour" particles, and $\gamma_{\mathrm{h4}}$ defines the floating point operations counts of $4^{\mathrm{th}}$ Hermite scheme per particle per interaction per step, from the work \citet{Nitadori+Makino+2008} which is a constant value has  $\gamma_{\mathrm{h4}}=60$. 

Bandwidth is measured as a part of hardware performance along with computing speed ($P$).
In \mynb, we defined bandwidth ($B_{\mathrm{reg}}, \ B_{\mathrm{irr}}$) as 
\begin{equation} \label{eqnbb}
B_{\mathrm{reg}} = \frac{N_{\mathrm{mcr}}}{T_{\mathrm{mcr}}} = \frac{8\cdot (41+l_{max}) \cdot N / N_p}{T_{\mathrm{mcr}}}
,\qquad
B_{\mathrm{irr}} = \frac{N_{\mathrm{mci}}}{T_{\mathrm{mci}}} = \frac{8\cdot 19 \cdot \langle N_{\mathrm{act}} \rangle / N_p}{T_{\mathrm{mci}}}
\end{equation}
where $N_{\mathrm{[mcr|mci]}}$ is the number of bytes transferred during MPI communication after regular/irregular blocks, constant terms derived from the size of datasets transferred, in which $l_{max}$ is the maximum size of neighbour lists set manually.
Detailed results of all these performance indicators presented in the next section.

\begin{figure}
   \centering
   \includegraphics[width=12.0cm, angle=0]{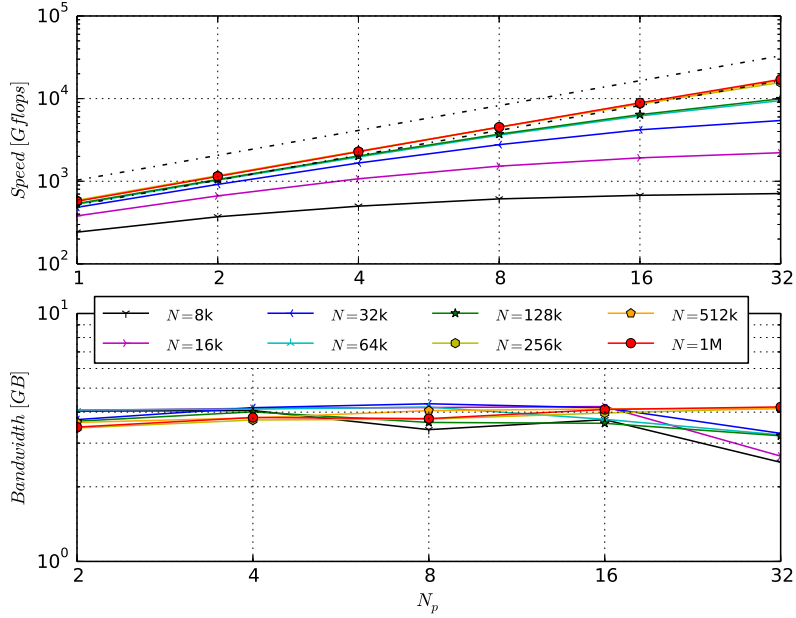}
   \caption{Hardware performance of \mynb running on ``MilkyWay" GPU cluster. The upper panel corresponds to the regular force computation speed ($P_{\mathrm{reg}}$), where two dashed lines refer to the peak single and double precision floating point performance. The lower panel corresponds to the bandwidth of regular part ($B_{\mathrm{reg}}$).}
   \label{Fig5}
\end{figure}

\subsubsection{Performance results}

The measured total wall-clock time of \mynb is shown in Figure~\ref{Fig1}.

On the whole, the result shows a good extensibility and acceleration when using more processors numbers. 
To be specific, we assign each part of the code with different weight. Among all of the parts the time spend for force computations always has the highest value, therefore both regular, irregular force computations and prediction have been implemented with parallel algorithm and decrease rapidly when code running in multi-processors.
Here in the heaviest part $T_{\mathrm{force}}$, the regular force computation $T_{\mathrm{reg}}$, which takes the highest fraction of computing time in the former code versions, has been accelerated and implemented on the specific device (GPU), as a consequence of causing a significant reduction of the whole running time costs. Other parts are currently executed on CPU side.

Table~\ref{table1} shows the Main components of \mynb as a function of $N$ and $N_p$. 
As describing above for every part expected scaling is evaluated by code structure, and fitting value is based on experimental data. The fitting process includes two steps by fitting $N$ and $N_p$ successively but independently. Firstly we used a minimum of fixed $N_p$ to avoid the disturbance of processor number and obtained the experimental scaling of $N$.
The fitting values with $N$ are obtained under circumstances which using an increasing particle numbers and fixed single processor ($N_p=1$), while for cases of multiprocessor-related value(i.e. $T_{\mathrm{mci}}$, $T_{\mathrm{mcr}}$, $T_{\mathrm{syn}}$, which have no numbers in single processor runs) the number of processors changed to $N_p=2$. 
The fitting values with $N_p$ are obtained by the second step. Grouping the dataset according to $N$, then dividing every group data by each $N$ dependent function to get the fitting values with $N_p$.
The fitting results of every main parts listed in the last column. 
Considering the expected scaling value of main parts, as $T_{\mathrm{mov}}$, $T_{\mathrm{syn}}$ and $T_{\mathrm{host}}$ have no significant and direct scaling with $N$ from code structure while $T_{\mathrm{pre}}$ makes up of two prediction branches that determined by $N$ in next time step, we expect these values as a simple exponent form of $N$. 
$N_{\mathrm{reg}}$, $N_{\mathrm{irr}}$ and $\langle N_{nb} \rangle$ which used in $T_{\mathrm{reg}}$ and $T_{\mathrm{irr}}$ are values which are completely dependent with $N$ as $N_{\mathrm{reg}} \propto N^{1.18}$, $N_{\mathrm{irr}} \propto N^{1.10}$ and $\langle N_{nb} \rangle \propto N^{2/3}$, then in the fitting values they are combined together also as an exponent form of $N$. 
At last, an unified exponent form of $N$ and $N_p$ are used in the last column rather than other symbols used in the middle column.

By taking the fitting values into the definition of experimental speed-up, we give the prediction about the performance of \mynb, which is shown in Figure~\ref{Fig2}. As a result, the optimal value of $N_p$ needed for larger simulations of different scales showed clearly from the figure.

Figure~\ref{Fig5} shows the hardware performance of \mynb in the actual environment on a real GPU accelerated cluster. Because GPUs played as the central role in acceleration we focus on GPU relevant parts in \mynb, $P_{\mathrm{reg}}$ and $B_{\mathrm{reg}}$ were drawn in the figure. 
For the figure of $P_{\mathrm{reg}}$, two dashed lines i.e. peak single and double precision floating point performance are used as the baseline, and the computation speed of different group of $N$ runs are increasingly closer to the peak when $N$ doubled. 
In the whole \mynb data structures there is two type precisions used in respective parts. 
Double precision type is used in main loops of the code which declared as the type ``REAL*8" in global header file, while for the regular force computation part which accelerated by GPU all of the data convert to the type ``REAL*4" then single precision is used in all relevant parts of CUDA routine. 
Therefore a mixing precision data structure is used in the regular part of \mynb - double precision in data moving process and single precision in data computation process. 
As the computation process in GPU card dominating the regular part, we use single precision to make a comparison.
As shown in the figure, the force computation speed of large $N$ runs passed over half of the maximum single precision performance (for instance, $P_{\mathrm{reg}}$ for 1M particle runs have the values $530 \sim 570$ Gflops per M2050/M2070 GPU card.)
This proportion concurs with the results of \citet{Berczik+etal+2011, Berczik+etal+2013}, which claimed the speed performance of another direct \nbody code \texttt{$\phi$-GPU} getting the values $\propto 550$ Gflops per C2050 GPU card and $\propto 1.48$ Tflops per K20 GPU card, both results approached half of the single precision performance peak.
Considering the hardware architecture, as operations among various register and memory causes extra inevitable time consumption, the proportion is acceptable in practical environments.
For the figure of $B_{\mathrm{reg}}$ ignoring the ``dropping" points, others remained with the level of more than $80\%$ of the maximum bandwidth performance.

\subsection{Performance of \mybs}
\label{sec:bonsaiptitle}

\subsubsection{Performance model}
\label{sec:bonsaimodel}

\begin{figure}
   \centering
   \includegraphics[width=12.0cm, angle=0]{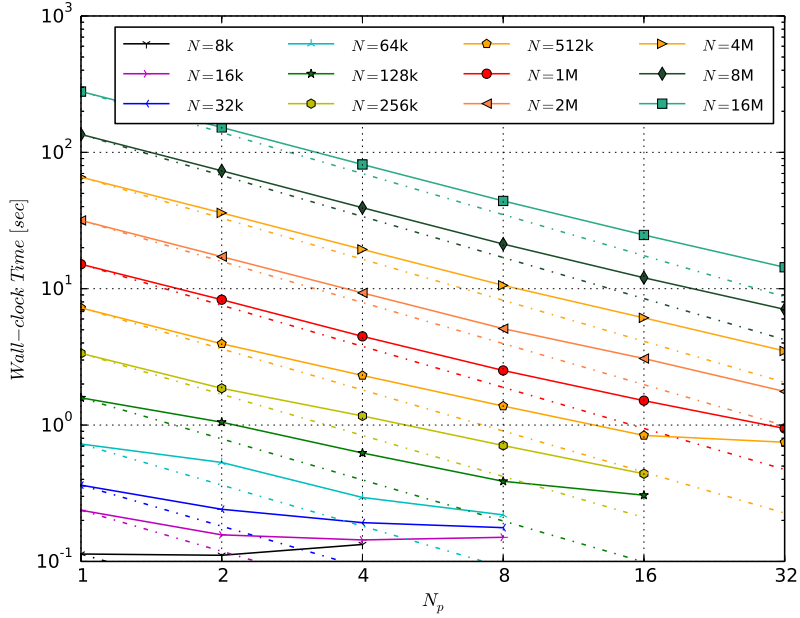}
   \caption{Total wall-clock time ($T_{\mathrm{tot}}$) of \mybs as a function of $N$ \& $N_p$. The legend is the same as Figure~\ref{Fig1}. Opening parameters of initial condition set as $\Delta t = 0.0625$, $\epsilon = 0.01$, $\theta = 0.5$.} 
   \label{Fig3}
\end{figure}

Similarly to Equation~\eqref{eqnbt}, in \mybs the total wall-clock time $T_{\mathrm{tot}}$ can be written as
\begin{equation}
  T_{\mathrm{tot}} = T_{\mathrm{tree}} + T_{\mathrm{force}} + T_{\mathrm{comm}} + T_{\mathrm{other}}
\end{equation}
where $T_{\mathrm{tree}} = T_{\mathrm{sort}} + T_{\mathrm{build}} + T_{\mathrm{prop}} + T_{\mathrm{grp}}$ is time spend on tree building, which mainly included sorting and reordering of the particles along a 1D number string mapped along \emph{Space Filling Curve}, tree structure construction, tree-node properties computation and active groups setting for next steps; 
$T_{\mathrm{force}}$ is time spend on force computations in tree-traverse; 
$T_{\mathrm{comm}} = T_{\mathrm{dom}} + T_{\mathrm{exch}} + T_{\mathrm{syn}}$ is time spend on distributing or redistributing the particles interprocessor and synchronizations of processors; 
$T_{\mathrm{other}} = T_{\mathrm{pre}} + T_{\mathrm{corr}} + T_{\mathrm{ene}}$ is time consumptions for other mainly essential parts, like local-tree predictions before tree construction, corrections after force computations and energy check. 
All of the time variables are measured by CUDA C function \texttt{cuEventElapsedTime} from CUDA Event Management Driver API and counted on device (GPU) entirely. Trivial time consumptions on the host side is ignored.
The entire descriptions are gathered in the last glossary (Table~\ref{table3}).

In hardware performance aspect, differed from Equation~\eqref{eqnbp} the force calculating speed in \mybs are written as
\begin{equation}
P_{\mathrm{force}} = \frac{N_{\mathrm{force\_tot}}}{T_{\mathrm{force}}} = \frac{N_{\mathrm{force}} \cdot \gamma_{\mathrm{t}}}{T_{\mathrm{force}}}
\end{equation} 
where $N_{\mathrm{force\_tot}}$ is the total floating point operation counts; $N_{\mathrm{force}}$ is the cumulative number of interactions; $\gamma_{\mathrm{t}}$ is the number of operation counts for each interaction in tree-code, which we used a constant value has $\gamma_{\mathrm{t}}=38$ from the work \citet{Warren+Salmon+1992, Kawai+etal+1999, Hamada+etal+2009, Hamada+Nitadori+2010}, and the result figure shows in Figure~\ref{Fig6}. Note that \citet{Bedorf+etal+2014} used another separated operation counts $23$ \&  $65$ for particle-particle and particle-cell interactions respectively.

\begin{table}
\bc
\begin{minipage}[]{100mm}
\caption[]{Main components of \mybs \label{table2}}
\end{minipage}
\setlength{\tabcolsep}{3.0pt}
\small
\newcommand{\myboxa}[1]{\raisebox{-2ex}[0pt][0pt]{\shortstack{#1}}}
\newcommand{\myboxb}[1]{\raisebox{-3ex}[0pt][0pt]{\shortstack{#1}}}
\begin{tabular}{ccccc}
\hline
\myboxa{Description} & \myboxb{Timing \\ variable} & \multicolumn{2}{c}{Expected scaling} & \myboxa{Fitting value [sec]} \\
 & & $N$ & $N_p$ & \\
\hline
Sorting and reordering & $T_{\mathrm{sort}}$ & $\mathcal{O}(N)$ & $\mathcal{O}(N_p^{-1})$ & $(1.5\cdot 10^{-6}\cdot N + 2.45\cdot 10^{-4}) \cdot N_p^{-1}$\\
Tree construction & $T_{\mathrm{build}}$ & $\mathcal{O}(N)$ & $\mathcal{O}(N_p^{-1})$ & $(2.8\cdot 10^{-7}\cdot N + 2.06\cdot 10^{-2}) \cdot N_p^{-1}$\\
Node properties  & $T_{\mathrm{prop}}$ & $\mathcal{O}(N)$ & $\mathcal{O}(N_p^{-1})$ & $(9.1\cdot 10^{-8}\cdot N + 5.78\cdot 10^{-3}) \cdot N_p^{-1}$\\
Set active groups & $T_{\mathrm{grp}}$ & $\mathcal{O}(N)$ & $\mathcal{O}(N_p^{-1})$ & $(1.7\cdot 10^{-9}\cdot N + 1.16\cdot 10^{-3}) \cdot N_p^{-1}$\\
Force computation & $T_{\mathrm{force}}$ & $\mathcal{O}(N\mathrm{log} N)$ & $\mathcal{O}(N_p^{-kp_{g1}})$ & $(2.5\cdot 10^{-6}\cdot N\mathrm{log} N - 0.10) \cdot N_p^{-0.88}$\\
Domain update & $T_{\mathrm{dom}}$ & $\mathcal{O}(N\mathrm{log} N)$ & $\mathcal{O}(1)$ & $5.4\cdot 10^{-10}\cdot N\mathrm{log} N + 2.96\cdot 10^{-3}$\\
Exchange & $T_{\mathrm{exch}}$ & $\mathcal{O}(N\mathrm{log} N)$ & $\mathcal{O}(1)$ & $2.1\cdot 10^{-9}\cdot N\mathrm{log} N + 1.16\cdot 10^{-2}$\\
Synchronization & $T_{\mathrm{syn}}$ & $\mathcal{O}(N^{kn_{s}})$ & $\mathcal{O}(kp_{s1}\cdot N_p^{kp_{s2}})$ & $(1.4\cdot 10^{-4}\cdot N^{0.45} + 9.3\cdot 10^{-4	})(0.5\cdot N_p^{0.49})$\\
Prediction & $T_{\mathrm{pre}}$ & $\mathcal{O}(N)$ & $\mathcal{O}(N_p^{-1})$ & $(1.5\cdot 10^{-8}\cdot N + 1.49\cdot 10^{-3}) \cdot N_p^{-1}$\\
Correction & $T_{\mathrm{corr}}$ & $\mathcal{O}(N)$ & $\mathcal{O}(N_p^{-1})$ & $(3.8\cdot 10^{-8}\cdot N + 7.88\cdot 10^{-4}) \cdot N_p^{-1}$\\
Energy check & $T_{\mathrm{ene}}$ & $\mathcal{O}(N)$ & $\mathcal{O}(N_p^{-1})$ & $(8.8\cdot 10^{-9}\cdot N + 7.14\cdot 10^{-4}) \cdot N_p^{-1}$\\
\hline
\end{tabular}
\ec
\tablecomments{0.8\textwidth}{Detailed descriptions of used symbol gathered in Table~\ref{table3}.}
\end{table}

\subsubsection{Opening parameters in tree-code}

Three opening parameters play significantly important roles in tree-code running and affect the performance consequently as a different result.

$\theta$: It is a dimension-less parameter defined in Equation~\eqref{eqmac} that controls the accuracy. 
Our test results showed that a smaller $\theta$ makes the running time increasing shapely, then stopped rising on a certain range ($\theta \approx 0.01$ as an experimental value); while a bigger $\theta$ ($\theta \geqslant 0.2 \sim 0.35$ influenced by $N$) causes a less accuracy of simulations. 

$\epsilon$: The softening parameter $\epsilon$ does not contribute to the running time, on the other hand an optimal $\epsilon$ could lead to a best approach to the minimum error. For a too small softening the estimates of the forces will be too noisy, while for a too large softening the force estimates will be misrepresented systematically, in between there is an optimal softening.
The optimal $\epsilon$ depends both on the number of particles and the size of time steps. From the work \citet{Athanassoula+etal+2000} when $\epsilon$ has different minimum values the conditions of simulations are not the same, there is a relationship between $N$ and optimal $\epsilon$. 
Through the comparison of their conclusions with groups of our test results of \mybs code (using $\Delta E$ instead of \emph{MASE}, $\theta=0.5$; $\Delta t = 0.0625$), we conclusion that the value of $\epsilon$ what leads to a minimum $\Delta E$ is consistent with conclusions of the reference.

$\Delta t$: The value of $\Delta t$ affect both running time and energy error.
On the running time side, $\Delta t$ has a noticeable linearly dependent with time as: $t \propto 1/\Delta t$; on the relative energy error side, the tests results are more complex. 
Our results shows that $\Delta E_{\mathrm{min}}$ varies sensitively under the chosen of combinations of $\Delta t$ and $\epsilon$, as well as the different $N$. 

\subsubsection{Performance results}

\begin{figure}
   \centering
   \includegraphics[width=12.0cm, angle=0]{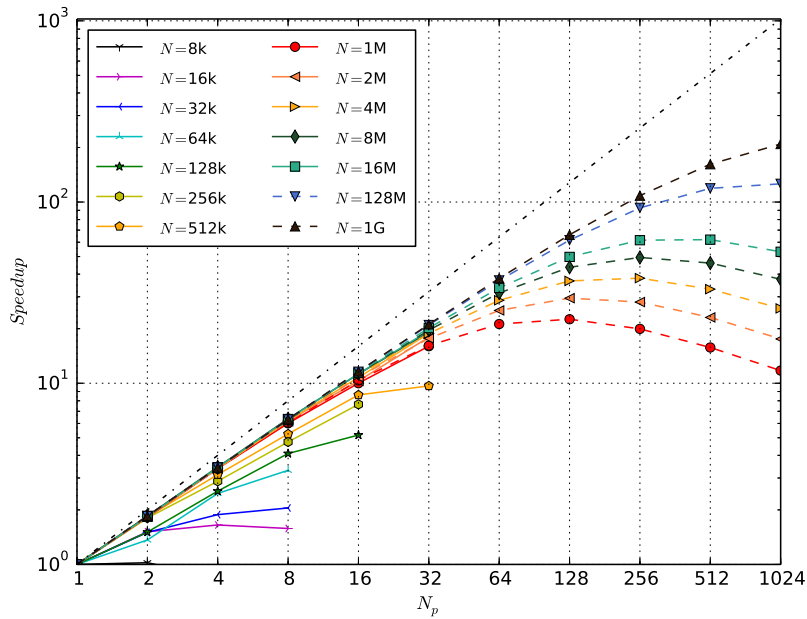}
   \caption{The speed-up ($S$) of \mybs as a function of particle number $N$ \& $N_p$. The legend is the same as Figure~\ref{Fig2}. Opening parameters of initial condition set as $\Delta t = 0.0625$, $\epsilon = 0.01$, $\theta = 0.5$.} 
   \label{Fig4}
\end{figure}

The measured total wall-clock time of \mybs is shown in Figure~\ref{Fig3}.

In Figure~\ref{Fig3} \& \ref{Fig4} we do not use any data for small particle number and large number of GPU nodes,
because the performance goes down and the GPU's are not fully loaded in this regime. 

We decomposed $T_{\mathrm{tot}}$ into main components as described in Section~\ref{sec:bonsaimodel}. 
For every components we measured the running time separately, and obtain fitting formulae as a function of
$N$ and $N_p$. The fitting procedures was the same as in the \mynb part described in Section~\ref{sec:nbody6++model}, the results are shown in Table~\ref{table2}.

Figure~\ref{Fig4} shows the experimental speed-up of \mybs defined as Equation~\eqref{eqse}. Compared with the result of Figure~\ref{Fig2}, \mybs has tendency to a lower peak but wider scope. Considering the weight factors of force computational part in Table~\ref{table1} \& \ref{table2} quantitative information of ascendant distinctness are revealed, while the different weight of communication part is the mainly determinant for descendant lines.

\begin{figure}
   \centering
   \includegraphics[width=12.0cm, angle=0]{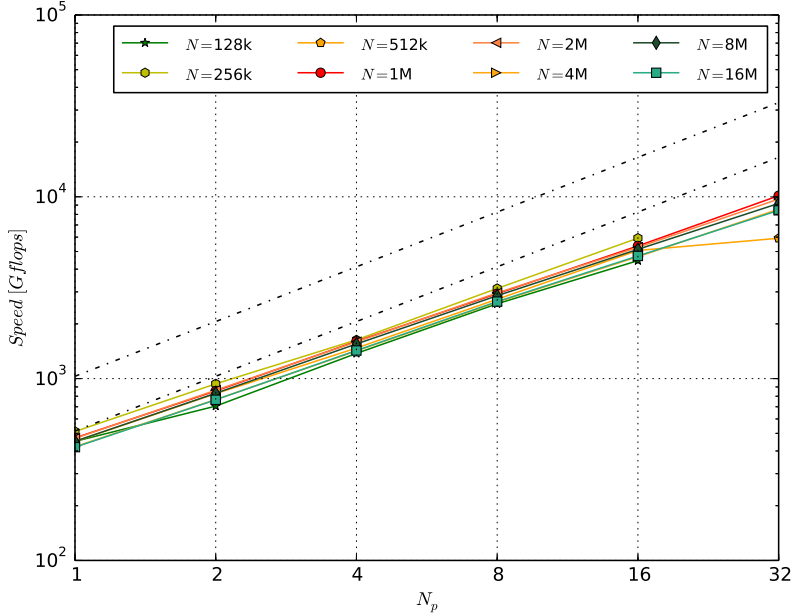}
   \caption{Hardware performance of force computation speed ($P_{\mathrm{force}}$) of \mybs running on ``MilkyWay" GPU cluster. Two dashed lines in the figure refer to the peak single and double precision floating point performance.}
   \label{Fig6}
\end{figure}

Figure~\ref{Fig6} shows the hardware performance of \mybs in a practical environment on a real GPU accelerated cluster. 
The performance in floating point operations per second is given for the dominant part of the force computations only.
The figure indicates that for the large enough $N$ we get near half of the peak single precision of our M2050/M2070 GPU card and increasing steadily for large scale $N_p$.
This proportion is similar to the result of \mynb code discussed in Section~\ref{sec:nbody6++ptitle}.
The utilization of the GPU is quite good considering the tree-code structure.

\section{DISCUSSION}
\label{sect:discussion}

\begin{figure}
   \centering
   \includegraphics[width=14.0cm, angle=0]{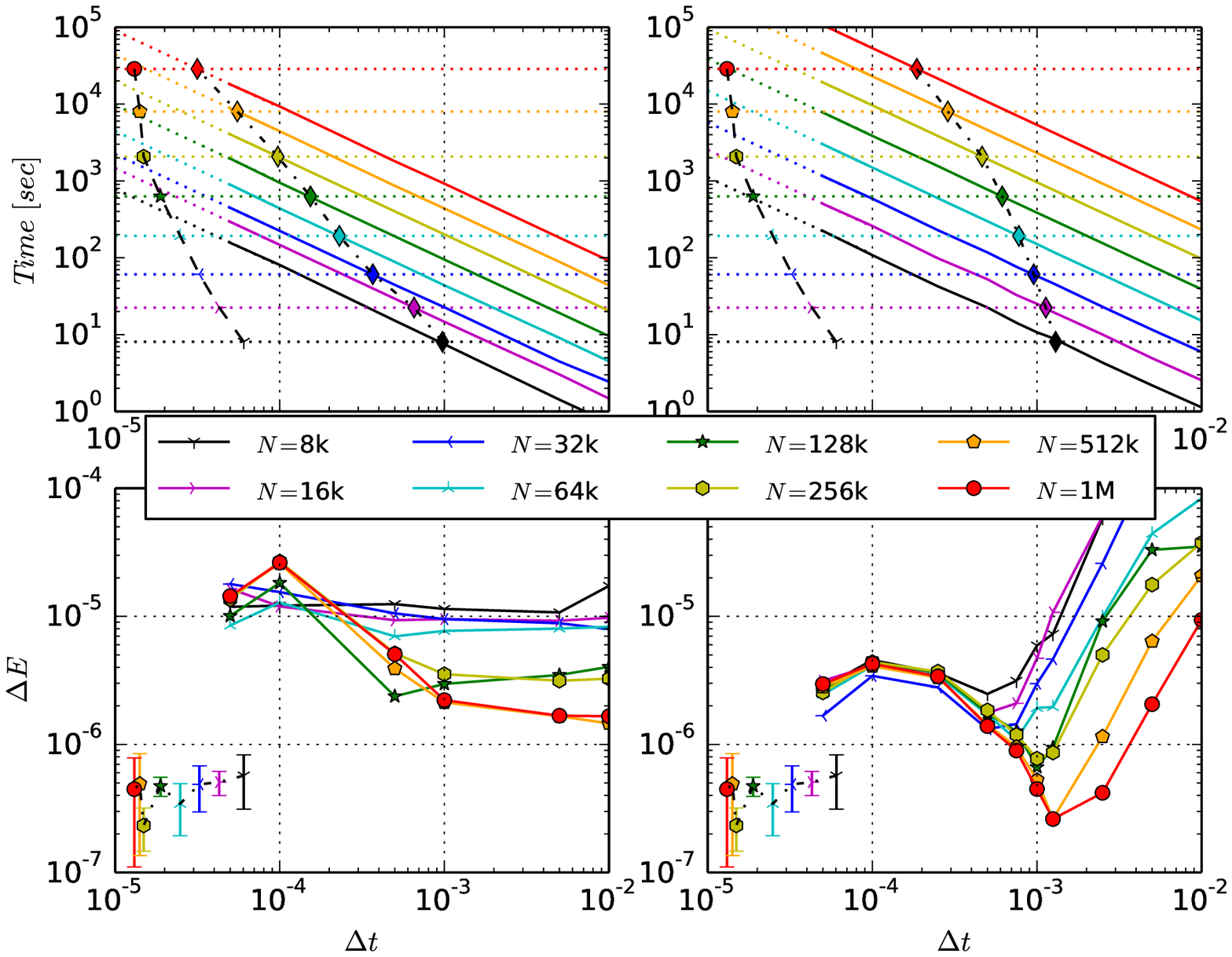}
   \caption{Comparisons of wall-clock time and relative energy error of \mynb and \mybs as a function of $\Delta t$. Opening parameters of \mybs set as $\epsilon = 0.01$, $\theta = 0.5$ in the left column, and smaller value $\epsilon = 0.001$, $\theta = 0.2$ as the control group in the right column. In every panel the left dashed line corresponds \mynb benchmark data, and solid lines are \mybs data. The diamond symbols indicates junctions of \mybs which has the same running time as \mynb in the case of the same $N$.}
   \label{Fig7}
\end{figure}

In this paper we analyze the performance of two very different kinds of \nbody codes, both pioneers in
their fields and both heavily optimized for GPU acceleration and parallelization - \mynb and \mybs.
There is always the question what is the turn-even point for the codes, how do they compare with each other.
Due to the very different nature of the two codes such a comparison is inevitably unfair - \mynb has few-body 
regularizations and is aimed for high
accuracy of both near and more distant gravitational forces; \mybs achieves optimal performance if the opening
parameter $\theta$ is relatively large, providing rather less accurate gravitational forces. 
But in certain ranges of parameters both codes may overlap in performance, accuracy and efficiency. It is the
goal of this paper to provide a quantitative information about this. 

We do this with the help of the four panels in Figure~\ref{Fig7} - they show wall-clock time and energy accuracy
as a function of the average time step; the main curves are for \mybs as indicated in the caption, for two
different opening parameters. However, also data for \mynb are shown for comparison: wall-clock time and 
accuracy as a function of average time step. In addition, we show that for a fixed particle number the time step of
\mybs which results to the same wall-clock time as for \mynb.

The following main conclusions can be drawn: at same wall-clock time and same particle number (and $\theta=0.2$)
\mybs runs typically with time steps of a factor of $10-50$ larger. In other words \mynb provides a much
smaller time step and a factor of $10$ better accuracy (see lower panels of Figure~\ref{Fig7}). 
Here energy error are used as the criterion to compare the accuracy. 
In our case the time evolution of the energy error contain two main parts. 
One part comes from the machine accuracy of the potential and force calculations.
This is in our cases close to the single precision machine accuracy of order $10^{-7} \sim 10^{-8}$. 
The other error component comes from the numerical integration process itself, which plays as the dominated role in total energy error.
In the \mynb we are using the complex Hermite $4^{th}$ order individual block time step integration combined with the Ahmad-Cohen neighbour scheme. We have chosen the time step parameter $\eta$ of the Aarseth time step criteria (for regular and irregular time step, which values set as 0.02 in initial input files) such that the energy error keeps the level of $10^{-6} \sim 10^{-7}$. How the global energy error of our integrator in \mynb behaves can be found in a comprehensive study by \citet{Makino+1991}.
In the case of the \mybs, the code using the simple leap-frog integration scheme, which (for the reasonable computational 
speed to reach the $1$ \textit{N}-body Time Unit) have a average energy error in a level of $10^{-5} \sim 10^{-6}$.
Insofar we have not discovered anything unexpected; however \mybs can reach surprisingly good accuracy in total energy
(like $5\cdot 10^{-6}$) at wall-clock times comparable to \mynb. With a larger opening angle ($\theta=0.5$)
the time step and wall-clock parameters approach each other more (factor two to three, for one million bodies).
In such a case still there is a quite considerable energy accuracy of order $10^{-5}$. That levels of energy
accuracy arguably may be sufficient even for collisional gravothermal systems, as it seems.

However, the total energy conservation is not the only criterion to judge about the use of a code and its
accuracy. In \mynb close encounters and interactions of compact or hierarchical multiple systems
are treated with regularization methods and zero softening, while \mybs uses an artificial softening of the
gravitational potential at small distances. Reasonable energy conservation refers to that artificial gravitational
potential including softening, which is conservative as well, but not the true few-body potential.
So the additional numerical efforts necessary for \mynb goes on
one hand side into the exact resolution of all kinds of close interactions below the softening length used in
\mybs. But also on the other side, the long-range interactions, \mybs uses the standard Tree-code procedure of
approximating forces from groups of particles by forces from their centers of masses and multipoles. This feature
needs to be tested by simulation of core collapsing star clusters, where long range gravitational interactions
determine the global evolution, which is beyond the scope of this paper.

\normalem
\begin{acknowledgements}
We want to thank \texttt{NBODY} series developer Sverre Aarseth for providing the \texttt{NBODY6} code and lectures about how to use it. 
We thank Long Wang for continuous developing the newest version of \mynb code. 
We also want to thank \mybs developers Jeroen B{\'e}dorf, Evghenii Gaburov and Simon Portegies Zwart for providing their \nbody code. 
We acknowledge support by Chinese Academy of Sciences through the Silk Road
Project at NAOC, through the Chinese Academy of Sciences Visiting Professorship
for Senior International Scientists, Grant Number 2009S1-5 (RS), and through the
``Qianren" special foreign experts program of China.

The special GPU accelerated supercomputer ``Laohu" at the Center of Information and
Computing at National Astronomical Observatories, Chinese Academy of Sciences, funded
by Ministry of Finance of People's Republic of China under the grant ZDY Z2008-2,
has been used for the simulations, as well as the supercomputer ``The Milky Way System" 
at J{\"u}lich Supercomputing Centre in Germany, built for SFB881 at University of Heidelberg,
Germany.

PB acknowledge the special support by the NAS Ukraine under the Main Astronomical
Observatory GPU/GRID computing cluster project.

\end{acknowledgements}

\bibliographystyle{raa}
\bibliography{bibtex}

\begin{table}
\bc
\begin{minipage}[]{100mm}
\caption[]{Glossary \label{table3}}
\end{minipage}
\small
\renewcommand{\arraystretch}{0.95}
\begin{tabular}{lp{30em}} \hline
Variable & Description \\
\hline
\multicolumn{2}{l}{\textsc{General}\parbox[l][4ex]{0pt}} \\
$N$ & total particle number \\
$N_p$ & number of processors \\
$S_a$ & theoretical maximum speed-up defined by Amdahl's law \\
$S_i$ & ideal maximum speed-up equals to $N_p$ \\
$S_e$ & experiential speed-up equals to the ratio between measured time of single and multiple processor number \\
$P$ & force computation speed of floating point operations per second \\
$B$ & bandwidth of bytes of data transfer per second \\
$T_{\mathrm{tot}}$ & total wall-clock time \\
$kn_{x}, kp_{x}$ & quantitative factors for fitting result of certain parts; $k[n|p]$ implies the factor only depends on $N|N_p$, subscript $x$ indicates different parts \\
$\Delta E$ & relative energy error \\
$\Delta t$ & time step interval of integration \\
\hline
\multicolumn{2}{l}{\texttt{NBODY6++}\parbox[l][4ex]{0pt}} \\
$\langle N_{\mathrm{act}}\rangle $ & average number of integrated active particles \\
$\langle N_{\mathrm{nb}}\rangle $ & average neighbour number \\
$N_{\mathrm{irr}}$ & cumulative number of irregular time steps \\
$N_{\mathrm{reg}}$ & cumulative number of regular time steps \\
$\gamma_{\mathrm{h4}}$ & floating point operations counts per particle per interaction per step \\
$T_{\mathrm{comm}}$ & sum of communication time \\
$T_{\mathrm{force}}$ & sum of force computation time \\
$T_{\mathrm{host}}$ &  time spend on the host side \\
$T_{\mathrm{irr}}$ & neighbour (irregular) force computation time \\
$T_{\mathrm{mci}}$ & MPI communication after irregular blocks \\
$T_{\mathrm{mcr}}$ & MPI communication after regular blocks \\
$T_{\mathrm{mov}}$ & time spend on data moving for parallel runs \\
$T_{\mathrm{pre}}$ & particle prediction time \\
$T_{\mathrm{reg}}$ & full (regular) force computation time \\
$T_{\mathrm{syn}}$ & interprocessor synchronization time \\
\hline
\multicolumn{2}{l}{\texttt{Bonsai}\parbox[l][4ex]{0pt}} \\
$N_{\mathrm{force}}$ & cumulative number of interactions \\
$\gamma_{\mathrm{t}}$ & floating point operations counts per particle per interaction per step \\
$T_{\mathrm{build}}$ & time spend on tree structure building \\
$T_{\mathrm{comm}}$ & sum of communication time \\
$T_{\mathrm{corr}}$ & particle correction time \\
$T_{\mathrm{dom}}$ & time spend on update of particle domain \\
$T_{\mathrm{ene}}$ & energy check time \\
$T_{\mathrm{exch}}$ & time spend on particle exchange \\
$T_{\mathrm{force}}$ & force computation time \\
$T_{\mathrm{grp}}$ & time spend on setting active groups\\
$T_{\mathrm{pre}}$ & local tree prediction time \\
$T_{\mathrm{prop}}$ & node properties computation time \\
$T_{\mathrm{sort}}$ & sorting and data-reordering time \\
$T_{\mathrm{syn}}$ & interprocessor synchronization time \\
$T_{\mathrm{tree}}$ & sum of the whole tree construction time \\
$\epsilon$ & softening to diminish the effect of graininess \\
$\theta$ & opening angle to control the accuracy \\
\noalign{\smallskip}\hline
\end{tabular}
\ec
\end{table}

\end{document}